\documentstyle[aps]{revtex}
\begin{document}
\draft
\title{Probabilistic Fragmentation and Effective Power Law}
\author{Matteo Marsili and Yi-Cheng Zhang}
\address{Institut de Physique Th\'eorique, 
Universit\'e de Fribourg, CH-1700}
\date{\today}
\maketitle
\begin{abstract}
A simple fragmentation model is introduced and analysed. We show that,
under very general conditions, an effective power law for the mass 
distribution arises with realistic exponent. This exponent has 
a universal limit, but in practice the effective exponent depends on 
the detailed breaking mechanism and the initial conditions. 
This dependence is in good agreement with experimental
results of fragmentation. 
\end{abstract}

One of the best known physical processes in Nature is 
fragmentation. From our daily experience we know that a material bulk 
under stress or shock will break into smaller pieces. Experiments show
generally that 
the number of fragments with a linear size larger than $r$ behaves like
\begin{equation}
N(r)\sim r^{-D}.
\label{Dfract}
\end{equation}
The exponent $D$, usually called fractal dimension, is found to 
lie in the range $D\approx 2\div 3$ for fragmentation of 3-dimensional 
objects\cite{3d}. 
A theoretical understanding of the statistical origin of eq. (\ref{Dfract})
is currently pursuited by many authors. The simplest models \cite{lognorm} 
predict a log-normal distribution, incompatible with observation.
More refined models with various assumptions about breaking mechanism
\cite{3d,models}, yield a power law behavior for $N(r)$. 
However it appears that theories producing a single universal exponent
may not account for the experimental range of $D$.
Recently quite large amount of numerical simulations 
\cite{herrmann} with rather realistic physical parameters such as 
stress, shear, neighborhood were able to show qualitatively correct 
power laws. 

In this Letter we develop an analytical model of 
fragmentation, {\em without} using specific breaking mechanism. 
We shall see that under very general, simple conditions an effective 
power law will result. 
A general theory is called for, since it is difficult to observe what 
is really happening {\em during} fragmentation. 
We imagine that fragmentation happens in
a hierarchical order---i.e. a large piece breaks into $n$ smaller ones and
each of the $n$ pieces may then break further. For simplicity we assume
$n=2$ for all $d$ and levels. Our numerical simulations show that 
other finite $n$ does not affect our conclusions.

At level $k$ of the hierarchy, we consider one object of volume $V$ and 
energy $E$, the only variables retained in our model.
$E$ is the total energy, including kinetic energy, elastic energy, etc.
We stress, however, that at our level of description this specification 
is not necessary. 
The only property of the energy we retain is that it is conserved. 
That is not exactly true in real fragmentation but it is a good
approximation for large masses because usually dissipation is 
an effect proportional to the area of the fracture produced.
We assume further that it is the energy density 
$E/V$ that decides whether an object breaks:
if $E/V$ exceeds a threshold, set to $1$ for all $k$,
the object breaks further, otherwise not.
If $E/V>1$, the object breaks into two pieces of energy
$\varepsilon$ and $E-\varepsilon$ and volume $\upsilon$ 
and $V-\upsilon$ respectively. 
The two resulting fragments
will or will not break in their turn according to their 
energy--volume ratios, i.e. if
\begin{equation}
x_1=\frac{\varepsilon}{\upsilon}>1\; ;\; \;x_2=\frac{E-
\varepsilon}{V-\upsilon}>1.
\label{x1x2}
\end{equation}
The above process is repeated for an arbitrary number $k$ of levels.
Note that $k$ is {\em not} necessarily 
proportional to time. The above variables are all for level $k$ and we 
have suppressed the sub-index $k$ for clarity. 
At level $k=0$, $E_0$ and $V_0$ are
given by the initial energy and volume of the system
($E_0/V_0>1$).

Let $q(\varepsilon,\upsilon|E)d\varepsilon 
d\upsilon$ be the probability that the energy and 
volume of an element are between $\varepsilon$ 
and $\varepsilon+d\varepsilon$ and $\upsilon$ and $\upsilon+d\upsilon$,
given that it results from the fragmentation of
an object of unit volume $V=1$ and energy $E>1$ (for our purposes we may
consider the volume V to be unity at any level, since only the ratio
$E/V$ matters). This distribution accounts for all the
informations of a detailed breaking mechanism.
We shall assume that $q(\varepsilon,\upsilon|E)=\frac{1}{E}\varphi(\upsilon)$
\cite{notaq}. This implies an
uniform distribution in energy, but arbitrary distribution in volume 
(for symmetry we require $\varphi(\upsilon)=\varphi(1-\upsilon)$).
This certainly is not the most general case, but it still
includes a large class of models. 

The fragments with energy density $x>1$ are called ``unstable'',
those with $x<1$ are called ``stable''. 
The nature of the cascade process is best illustrated by
studying the distribution $p_k(x)$ of energy density
$x$ of the {\em unstable} elements at the level $k$.
This can be computed from the initial distribution $p_0(x)$
once a recurrence relation between $p_k(x)$ and
$p_{k+1}(x)$ is found. In order to derive this 
relation, let us consider a unit volume 
object with energy $E>1$ breaking in two smaller elements.
The jacobian of the transformation
$(\varepsilon,\upsilon)\to (x_1,x_2)$ 
is readily found using (2), and we find the 
joint distribution of $x_1$ and $x_2$ from 
the above  $q(\varepsilon,\upsilon|E)$, 
\begin{equation}
p(x_1,x_2)=\frac{1}{E}\varphi\left(\frac{E-x_2}{x_1-x_2}\right)
\left|\frac{(x_1-E)(x_2-E)}{(x_1-x_2)^3}\right|
\label{eq1}
\end{equation}
Denote $x_+=\max(x_1,x_2)$ and $x_-=\min(x_1,x_2)$,
from (2) we have $x_-\le E\le x_+$. Integrating
out one of the two variables we obtain the distributions of $x_\pm$, 
respectively:
\begin{eqnarray}
p_+(x|E)&=&\frac{2\theta(x-E)}{E}\int_0^{E/x}
v\varphi(\upsilon)dv\nonumber\\
p_-(x|E)&=&\frac{\theta(E-x)}{E}
\label{ppm}
\end{eqnarray}
where $\theta(x)=0$ for $x<0$ and $\theta(x)=1$ otherwise.
The $+$ element always breaks further since $x_+\ge E>1$, while
the $-$ element does not break if $x_-<1$, this occurs
with the probability 
\begin{equation}
C(E)=\int_0^1 p_-(x|E)dx=\frac{1}{E}.
\label{ce}
\end{equation}

Let us define the distribution of the energy density 
$p(x|E)$ of one {\em unstable} ($x>1$) element between
the two fragments.
The pair $x_-$, $x_+$ falls into two cases:
i) If $x_-<1$, which occur with probability
$C(E)$ given above, the only possibility is that $x=x_+$.
In this case the distribution $p(x|E)$
is just that of $x_+$, i.e. $p_+(x|E)$. 
ii) If $x_->1$, which occur with probability $1-C(E)$, $x$  can 
be either $x_+$ or $x_-$. In this case, occurring with the
probability $1-C(E)$, $p(x|E)$ is the average of
the distributions $p_\pm(x|E)$ with weight $1/2$ each.
Taking into account the two cases we find:
\[p(x|E)=\theta(x-E)\frac{E+1}{E^2}\int_0^{E/x}
\upsilon \varphi(\upsilon)d\upsilon+\theta(E-x)\frac{1}{2E}.\]
One can verify that $p(x|E)$ is normalized in $[1,\infty)$.
Using $p(x|E)$ we finally construct the iteration relation:
\begin{equation}
\label{iter}
p_{k+1}(x)=\int_1^\infty \!dE p(x|E)p_{k}(E).
\end{equation}
  
As $k\to\infty$ the distribution $p_k(x)$ is expected to 
converge to a limit which describes the asymptotic behavior:
\begin{equation}
p_\infty(x)= \int_1^\infty p(x|E) p_\infty(E)dE.
\label{pinf}
\end{equation}
The solution of this equation will depend on the specific
choice of $\varphi(\upsilon)$. For the special case 
$\varphi(\upsilon)=1$, i.e. the uniform distribution which is interesting 
{\em per se}, eq. (\ref{pinf}) can be solved to give:
$p_\infty(x) = \frac{A}{x}\left(1-\frac{1}{2x}\right)
\exp\left(-\frac{1}{2x}\right)$.
This distribution is not normalizable because $p_\infty(x)\sim A/x$ for 
large $x$. For arbitrary $\varphi(\upsilon)$, 
detailed study of eq. (\ref{pinf}) for $x\to\infty$
reveals that this asymptotic behavior holds as well.
Indeed setting $p_\infty(x)\sim A x^{-\gamma}$ in 
eq. (\ref{pinf}) and carrying out an asymptotic analysis, one finds 
that the exponent $\gamma$ must satisfy the equation
\begin{equation}
\gamma = \int_0^1\upsilon^{1-\gamma} \varphi(\upsilon)d\upsilon.
\label{gamma}
\end{equation}
This equation has always a solution for $\gamma=1$. 
For convex distributions $\varphi(\upsilon)$, i.e. they weight more the center
of the interval $[0,1]$, there is a second solution $\gamma>1$.
However the solution with $\gamma=1$ clearly dominates 
the large $x$ behavior.
For concave $\varphi(\upsilon)$ the second solution becomes $\gamma<1$. 
We expect the physical relevant distributions are convex since objects
tend to break in the middle easier than at the edges.

Denote by $C_k$ the probability that one of the 
fragments (the other is unstable by definition) becomes {\em stable} 
at level $k$:
\begin{equation}
C_k=\int_1^\infty C(E)p_k(E)dE .
\label{ck}
\end{equation}
The important consequence of $\gamma=1$ is that $C_k\to 0$ as 
$k\to\infty$, regardless of the distribution 
$\varphi(\upsilon)$.
To see this, we assume for simplicity that $p_k(x)= 
A_k x^{-\gamma_k}$ with $\gamma_k\to 1^+$  as $k\to\infty$. 
Then the integral in (\ref{ck}) is finite
for all $\gamma_k>1$. On the other hand, as $\gamma_k\to 1^+$,
the constant $A_k$ which enforces normalization vanishes.
Therefore $C_k\sim A_k$ vanishes as well.

Unfortunately we cannot obtain $C_k$ in closed form even for the uniform 
distribution $\varphi=1$. The numerical
procedure is straightforward: start with a distribution at level $k=0$ with
an initial condition $p_0(E)=\delta (E-E_0), E_0>1$, iterate eq. (6)
to the desired level $k$, then find $C_k$ using (9). 
The numerical results for a particular $\varphi$ and various initial 
energies are plotted in Fig. \ref{figck}. 
The function $C_k$ is not universal. It depends
on $E_0$ as well as on the function $\varphi$.
In general, for a given $\varphi$, $C_k(E_0)$
decreases with $E_0$ for fixed $k$, and vanishes
slowly when $k\to \infty$ (see fig. \ref{figck}).

From the above preparations, 
we can establish the desired scaling laws.
In eq. (\ref{Dfract}) only the {\em stable} 
fragments left in the cascade count. 
$N(r)$ is trivially related to 
the distribution $W(V)$ of volumes $V=r^d$: 
$N(r)=\int_{V}^\infty W(V)dV$
where $W(V)dV$ is the number of stable 
fragments with volume between $V$ and $V+dV$.
We thus have the relation from (1):
\begin{equation}
W(V)\propto V^{-\alpha-1},\;\;\;\;\;\; D=d\alpha.
\label{ansatz}
\end{equation}
$W(V)$ receives contributions $W_k(V)$ from all fragmentation 
levels and it can be written:
\begin{equation} 
W(V)=\sum_{k=1}^{\infty}W_k(V)\simeq\sum_{k=1}^{\infty} 
C_k N_{k} w_k(V).
\label{W(V)}
\end{equation}
Here $w_k(V)$ is the volume distribution of all the fragments (stable
and unstable) produced by the $k^{\rm th}$ step of fragmentation, $N_k$ is the
number of {\em unstable} fragments at level $k$. $N_k$ unstable objects 
produce $2N_k$ fragments, of which $(2-C_k)N_k$, which is $N_{k+1}$, are 
unstable again. Thus $C_k N_{k}$ {\em stable} fragments are produced at 
level $k$. Eq. (\ref{W(V)}) assumes that stable fragments are produced 
at level $k$ with a probability $C_k$ which is independent of their
volume $V$. This is clearly an approximation because in reality
$C_k=C_k(V)$ \cite{notaupp}. Neglecting this dependence allows us to keep
the discussion at an elementary level. Moreover for distributions 
$\varphi(\upsilon)$ 
which are sharply peaked around $\upsilon=1/2$ this approximation
is reliable. Indeed for $\varphi(\upsilon)=\delta(\upsilon-1/2)$ 
the dependence on the volume
clearly drops (all the fragments at level $k$ have the same volume). 
For general distributions this approximation yields an upper bound
to the true exponent $\alpha$\cite{notaupp}.
We shall see below that even for broad distributions, such as the uniform
one, we recover qualitatively correct results. As we shall see the fractal 
dimension is determined by the exponential behavior of $N_k$ with $k$, which
depends on $C_k$, and the explicit factor $C_k$ in eq. (\ref{W(V)}) does
dot play any role. This observation supports the present approximation.
Moreover the same qualitative behavior discussed below results from
a more detailed calculation, to be presented elsewhere\cite{elsewhere}.

For the uniform case $\varphi(\upsilon)=1$ it is easy to find \cite{nota}
that $w_k(V)=(\ln V)^{k-1}/(k-1)!$. Assuming
$C_kN_{k}\propto (2-C^*)^k$, one can easily sum eq. 
(\ref{W(V)}) with the result $W(V)\propto V^{-\alpha-1}$ with
$\alpha =1-C^*$. Let us generalize this  
analysis for an arbitrary distribution $\varphi(\upsilon)$.
Let us evaluate the $m^{\rm th}$ moment of $V$ using (\ref{W(V)}) and 
(\ref{ansatz}). Note that some moments can be divergent and there is 
a smallest $m^*$ for this to happen.
Multiply both sides of (\ref{W(V)}) by $V^m$ and integrate over $V$.
On the RHS one finds $\langle{V^m }\rangle_k$ where the average is done using
the distribution $w_k(V)$. Since $V$, within our approximation, is the 
product of $k$ independent variables,
each distributed by $\varphi$, we have 
$\langle{V^m}\rangle_k=\left[\int
\upsilon^m\varphi(\upsilon)d\upsilon\right]^k$. 
Multiplying $\langle{V^m}\rangle_k$ by $C_k N_{k}$
and summing over $k$, we see (e.g. by the ratio method)
that the sum first diverges when $(2-C^*)\int
\upsilon^{m^*}\varphi(\upsilon)d\upsilon=1$,
where $C^*$ is defined by $N_{k+1}=(2-C^*)N_k$.
This divergence must be matched to the one occurring on the 
LHS of the equation, which is proportional to
$\int dV V^{m-\alpha-1}$. The latter occurs for $m^*=\alpha$,
which therefore gives the relation
\begin{equation}
\int_0^1 \upsilon^{\alpha}\varphi(\upsilon)d\upsilon=\frac{1}{2-C^*},
\label{alpha}
\end{equation}
which implicitly yields the exponent $\alpha$ within the
present approximation. For
$\varphi=1$, eq. (\ref{alpha}) reduces to the result we found
previously. Most importantly, for $C^*=0$, and
only for this value, one finds $\alpha=1$, regardless of $\varphi$!

In order to derive the value of $C^*$, which determines the 
scaling exponent, we observe that in real life the observation 
of $W(V)$ is limited to a finite window $V_{\min}<V<V_0$, 
which may cover several decades. $C_k$ is a slowly 
varying function of $k$ whereas the typical $V$ decreases 
exponentially with $k$. Thus, over an exponentially large range 
of $V$, $C_k$ can be regarded as constant and an 
{\em effective} power law can be established.
More precisely, $W(V)$ in this window, is dominated
by contributions around a certain $k^*$ (see fig. \ref{figck})
thus giving $C^*=C_{k^*}$. Further fragmentation 
for $k>k^*$, seldom adds stable fragments larger than
$V_{\min}$. Pursuing this arguments further, one can find that,
to leading order, $k^*\simeq c \log (V_0/V_{\min})$
where $c$ is a coefficient of order unity
($c=1+\alpha$ for the uniform case within our approximation).
$k^*$ appears to depend weakly on the
initial energy $E_0$, as shown in Fig.1. 
Therefore, strictly speaking, there is no {\em true} power law,
except at the ideal limit $k^*\to\infty$ where one can take 
$C^*=0$ so that $D=d$. 

How does the exponent $\alpha$ vary in practical situations?
First we notice that often, in experiments, $V_{\min}$ is
set by a dissipation scale, below which an object cannot
break anymore. In other words at this scale the energy lost
in the rupture of an object becomes non-negligible with respect to 
its energy. Thus we conclude that $\alpha$ depends
on $V_0$ through the dependence of $C^*=C_{k^*}$ on 
$k^*\sim\log (V_0/V_{\min})$. This suggests that for 
$V_0/V_{\min}\to \infty$ it is possible to
recover the ideal limit $D=d$.
Moreover $C^*$ depends also on $E_0$.
A larger $E_0$ results in a smaller $C^*$, since 
$C_k$ decreases with $E_0$ (see fig. \ref{figck}).
Thus the effective $\alpha(E_0)$ increases with $E_0$, until
it reaches the universal value of $1$. 

Figure \ref{figalpha} shows that, in spite of the approximation
used to derive eq. \ref{alpha}, this scenario is confirmed
by numerical simulations. It also 
has many features that have been observed in real experiments.
In fragmentation of $2$ dimensional glass plates\cite{2d},
the fractal dimension was nicely extrapolated to 
$D=2$ for infinite input energy. 
The same agreement can be found comparing our results with $3$ 
dimensional data\cite{3d}.
We note in particular that $D$ for stony meteorites
and asteroids, which are the result of an extremely
long fragmentation process, where very high energies 
are involved, are in excellent agreement with $D=3$. 
For these systems we expect $V_0/V_{\min}$ and $E_0/V_0$ to be 
extremely large.
On the other hand, in projectile fragmentation
the input energy is the kinetic energy of the projectile.
The volume to be considered is the total volume of the system,
which is essentially that of the target object.
Therefore, even for extremely high kinetic energies, the 
input energy density $E_0/V_0$ which would enter our calculation 
can be relatively small. Indeed, experiments
of projectile fragmentation yield exponents $D\approx 2.5$, 
below the universal value $D=3$. 
Our analysis yields $d$ as an upper bound for 
$D$ ($C^*\ge 0$ implies $\alpha\le 1$). It is reassuring that,
apart from the data concerning ash and pumice, 
materials which probably need a separate treatment, all the 
data in ref.\cite{3d} have $D\le d=3$.  
Comparing the exponents obtained for different materials,
in terms of our model, translates into computing the 
exponent for different distributions $\varphi(\upsilon)$.
As shown in Fig. \ref{figalpha}, for $convex$ distributions the convergence
of $\alpha(E)$ to $1$ is faster than that obtained 
for {\em concave} distributions. 
This might be related to the fact that, from eq. (\ref{alpha}),
$\alpha\simeq 1-C^*/|4\langle \upsilon\ln \upsilon\rangle |+
O({C^*}^2)$, where the coefficient of $C^*$ is larger for
broader distributions.

Problems arise when considering the $d=1$ experiments 
described e.g. in ref. \cite{1d}. There it was found that
long thin glass rods fragmentation produces a size
distribution with an exponent $D\approx 1.5$.
This is clearly inconsistent with our analysis since it 
would need a $C^*<0$. This failure, we believe, lies in the
assumption that the breaking of a large object is determined
by its global energy density. Without loss of 
generality we know that energy correlation inside a volume propagates 
via nearest neighbor interaction, this leads to a Laplace equation. 
For correlation induced by a Laplace equation we know that for 
$d>2$ the correlation is very strong and $d=2$ 
is the marginal case. This strong correlation allows to
describe the fragmentation of one object as an event which
produces two objects and which depends on a single variable, 
its energy density. 
For $d=1$ this is not true. The energy is very loosely correlated along a 
line. We suspect that, because of the weak correlation, simultaneous 
breaking will happen in many uncorrelated regions of a large $d=1$ object, 
making our scenario invalid. 
For smaller and smaller rod length, the energy correlation becomes
stronger and stronger. We therefore expect that below
a certain length threshold, our scenario can be applied. Remarkably, in 
experiments of fragmentation of long glass rods \cite{1d}, a crossover 
occurs and for intermediate sizes the mass distribution is described 
by an exponent $D\approx 0.6<1$.

In this work we have analysed a simple fragmentation model. 
We show that under very general conditions an effective power law arises. 
The exponent is not universal but depends on the detailed mechanism
and the initial conditions. There is an ideal universal limit, independent
of any of our choices, which can be approached for higher input 
energies. This qualitative behavior, derived within a simple 
approximation for the mass distribution, is completely consistent 
with numerical simulations of our model. Furthermore our results agree 
reasonably well with existing experiments in 2-$d$ and 3-$d$. 
We also argue that our model is not directly applicable in $d=1$.
A more precise quantitative theory, assuming a general distribution
$q(\varepsilon,\upsilon)$, going beyond the approximations in 
eq. (\ref{W(V)}) and including explicitly the effects of dissipation,
is under current investigation \cite{elsewhere}.

This work was supported by the Swiss National Foundation under
grant 20-40672.94/1.

\begin{figure}
\caption{$C_k$ as a function of $k$ for $V_0=1$ and 
$E_0=2$ ($\Diamond$), $E_0=8$ ($\Box$) and $E_0=32$ ($\circ$). 
The data refers to the tent distribution $\varphi(\upsilon)=
2\upsilon$ for $\upsilon<1/2$ and $\varphi(1-\upsilon)=\varphi(\upsilon)$. 
The lines refers to
the contribution $\int_{V_{\min}}^1 W_k (V) dV$
of the level $k$ to the statistics 
of $W(V)$ for $V_{\min}<V<1$ with $V_{\min}=3\cdot 10^{-5}$.
These lines refer, from left to right, to the same values
of $E_0=2,8$ and $32$ used for $C_k$.}
\label{figck}
\end{figure}

\begin{figure}
\caption{Fit of the mass size distribution for $\varphi(\upsilon)=1$
and $E_0/V_0=4$ ($\Box$) and for the tent distribution 
and $E_0/V_0=8$ ($\Diamond$) and for $\varphi(\upsilon)=\delta(v-1/2)$
and $E_0=16$ ($\circ$). Inset: exponent $\alpha$, obtained from the fit, versus
$E_0/V_0$ for the uniform ($\Box$), the tent ($\Diamond$) and the delta 
($\circ$) distributions.}
\label{figalpha}
\end{figure}


\begin{thebibliography}{99}
\bibitem{notaq} It can be shown that this choice of $q$ results
from requiring the ``scaling form'' $q(\varepsilon,\upsilon)=
g(\varepsilon/\upsilon|E) \varphi(\upsilon)$, 
which depends on energies only through their densities.
\bibitem{notaupp} This results from observing that in general $C_k(V)$
should be an increasing function of $V$: larger fragments have larger 
probability of being stable than smaller ones because their energy is
assigned independently.
\bibitem{nota} Indeed the volume of an element at level $k$ is 
$V=\prod_{j=1}^k \upsilon_j$ where $\upsilon_j$ are uniform variables.
Then $-\ln V$ is a sum of $k$ exponential random variables and
it has a Gamma distribution with index $k-1$.
\end{thebibliography}
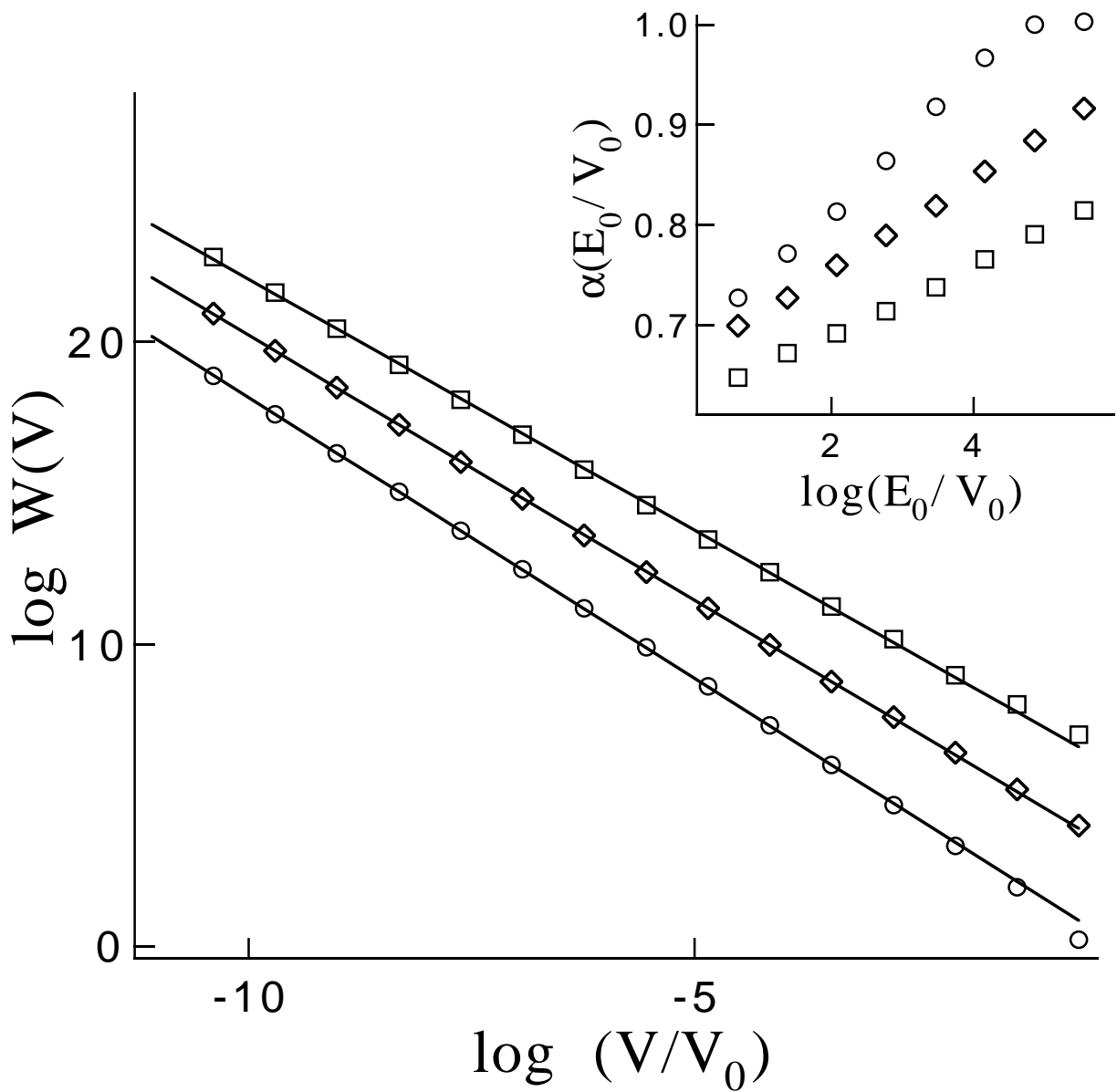
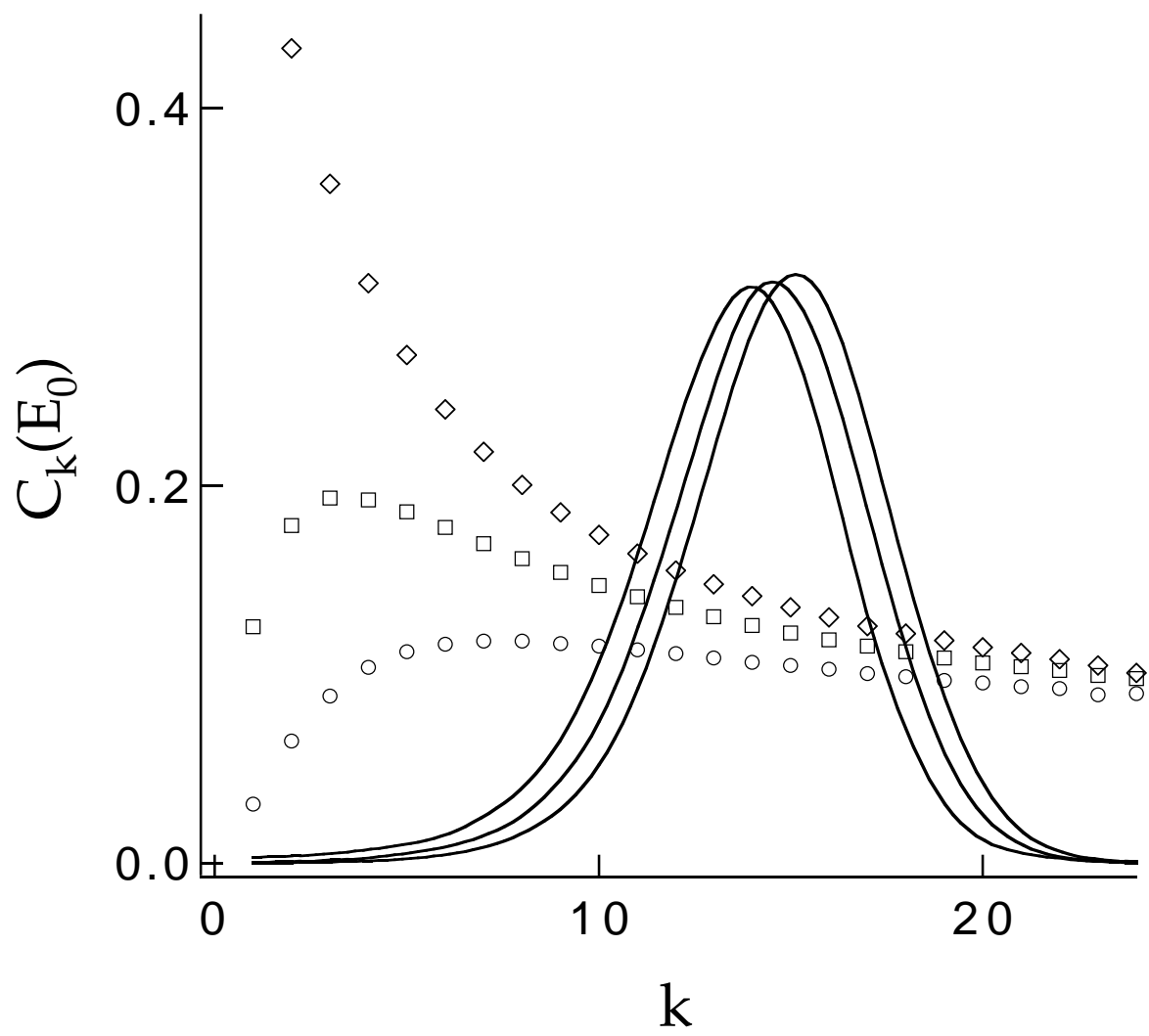
\end{document}